\documentclass[aps,pre,twocolumn,showpacs,floatfix]{revtex4}

\usepackage{url}
\usepackage{graphicx}
\usepackage{amsmath,psfrag,amssymb}
\usepackage{booktabs}
\usepackage{rotating}
\usepackage{tabularx}
\usepackage{subfigure}

\begin{document}

\title{Finding attractors in asynchronous Boolean dynamics} 

\author{Thomas Skodawessely}
\affiliation{Bioinformatics Group, Institute for Computer Science,
Leipzig University, H\"artelstrasse 16-18, 04107 Leipzig, Germany}

\author{Konstantin Klemm} 
\affiliation{Bioinformatics Group, Institute for Computer Science,
Leipzig University, H\"artelstrasse 16-18, 04107 Leipzig, Germany}

\date{\today}

\begin{abstract}
We present a computational method for finding attractors
(ergodic sets of states) of Boolean networks under
asynchronous update. The approach is based on a
systematic removal of state transitions to render the
state transition graph acyclic. In this reduced state
transition graph, all attractors are fixed points that
can be enumerated with little effort in most instances.
This attractor set is then extended to the attractor set
of the original dynamics. Our numerical tests on
standard Kauffman networks indicate that the method is
efficient in the sense that the total number of state
vectors visited grows moderately with the number of
states contained in attractors.
\end{abstract}

\maketitle

\section{Introduction} \label{sec:intro}

Complex disordered systems with many degrees of freedom can often be
approximated by {\em Boolean} dynamics \cite{Drossel2008,Bornholdt2005}. In
particular, gene-regulatory systems in living cells \cite{deJong:2002} have
been  modeled by Boolean dynamics since the 1960s \cite{Kauffman1969}. Each gene
is represented by a node with two possible states. In this {\em coarse-grained}
state representation, only high (Boolean true $=1$) and low (Boolean false $=0$)
chemical concentration of the gene product are distinguished. The regulatory
biochemical interactions between gene product concentrations are captured as
logical rules in the Boolean model. Each unit is assigned a Boolean function
(truth table) according to which it updates its state based on the states of
other units. In recent years, such Boolean models have been shown to capture
the dynamics of real regulatory systems
\cite{Albert2003,Li2004,Davidich2008,MacLean2010}.

The long-term behaviour of Boolean dynamics is of particular interest. It is
characterized by {\em attractors} (ergodic sets) as minimal subsets of the state
set from which the dynamics does not escape.  Attractors in a Boolean system
have been interpreted as distinct cell types in multicellular organisms
\cite{Kauffman1993,Ribeiro2007}. The computational problem of finding all
attractors in a Boolean system is difficult. Even the simpler problem of
deciding whether the system has a fixed point (the smallest possible attractor)
is NP-complete \cite{Milano2000,Garey:1979a}. In many instances of Boolean
networks, however, the state space to be searched may be largely reduced
\cite{Bilke2001,Richardson2005}, allowing for attractor detection in sparse networks with
several dozens of nodes. Known methods
\cite{Irons:2006,zhang_algorithms_2007,Dubrova2010} are tailored for dynamics under {\em deterministic synchronous}
update. The assumption of fully deterministic operation of all nodes in the
network may not be justified when modeling real regulatory systems. In fact,
the number and size of attractors change dramatically when giving up
deterministic synchronous update \cite{Samuelsson2003} and using stochastic
asynchronous update instead \cite{Greil:2005}.

Here we present a generally applicable exact method for attractor search in
Boolean dynamics under asynchronous single-node update. In non-rigorous terms,
the method works as follows. Departing from the asynchronous Boolean dynamics to
be analyzed, we modify the Boolean functions of a subset $X$ of the nodes such
that they cannot leave a ``preferred'' state (0 or 1) once they have reached it.
This reduction of the allowed state transitions is guaranteed not to eliminate
any of the original attractors: While additional attractors may appear, the
existing ones may lose some of their states but never disappear. In fact, we can
choose the set $X$ of nodes such that all attractors lose all but one state,
i.e.\ they become fixed points. This is the case when each directed cycle of the
Boolean network contains at least one vertex in $X$. In other words, removal of
the nodes in a so-called {\em feedback vertex set} $X$ leaves the Boolean
network acyclic. On sparse networks, feedback vertex sets $X$ can be found with
a cardinality $|X|$ much smaller than that of the whole node set. Then the
number of attractors (all fixed points) is not greater than $2^{|X|}$. These
attractors can be found by systematic enumeration. Now by construction, each
attractor of the original dynamics must contain at least one fixed point found
for the reduced dynamics. Finally the attractors are found by depth first
searches seeded at each fixed point of the reduced dynamics.

The rigorous description of this method in section~\ref{sec:method} uses formal
notions of Boolean mappings, operators, attractors and directed graphs. These
notions and their relations are introduced in section~\ref{sec:defs}.
Results and concluding remarks are presented in sections
~\ref{sec:results} and~\ref{sec:con}.

\section{Technical background} \label{sec:defs}

\subsection{Boolean dynamics and update operators}
A Boolean dynamical system of $n$ units (or {\em nodes}) is defined by
assigning each node $i \in \{1,\dots,n\}=:n_\rfloor$ a Boolean function
\begin{equation}
f_i: B^n \rightarrow B
\end{equation}
where $B=\{0,1\}$ is the set of elementary Boolean states.  The interaction
network underlying this system is extracted from the functions $f_i$. There is
an edge from node $j$ to node $i$ if and only if function $f_i$ explicitly
depends on the $j$-th coordinate. Put differently, $(j,i)$ is an edge if there
are state vectors $x,y \in B^n$ differing only at coordinate $j$ such that
$f_i(x) \neq f_i(y)$.  

The time-discrete dynamics of the system is made precise by defining the {\em
update mode} where {\em synchronous} ({\em parallel}) update is often used. Then
each node computes the state at time $t+1$ by applying its Boolean function
to the state vector at time $t$. A common alternative is the fully {\em
asynchronous} ({\em serial}) update mode. At each time step, only one node $u(t)
\in \{1,\dots,n\}$ is updated while all others keep their state. As a
generalization, a set of nodes potentially containing more than one but not
all nodes may be chosen individually at each time $t$. 

We formalize the update modes by defining an update operator $U_I$ for
each $I \subseteq \{1,\dots,n\}$. The operator affects a Boolean state
vector $x$ according to
\begin{equation}
(U_I x)_i = \left\{ \begin{array}{ll}
f_i (x), & \text{if } i \in I\\
x_i,     & \text{otherwise} 
\end{array}\right.
\end{equation}
Then the dynamics is given by the set $\mathcal{U}$ of such update operators
one of which is chosen at each time step for updating the state vector.  
For synchronous update, we have $\mathcal{U} = \{ U_{\{1,\dots,n\}} \}$ because the
only allowed update involves all nodes. Asynchronous update for single nodes is
performed with the operator set $\mathcal{U} = \{ U_{\{i\}} : i \in \{1,\dots,n\}
\}$. Operators may be applied with different probabilities. However, we do
not deal with probabilities here because the attractors of the system do not
depend on them as long as each operator in $\mathcal{U}$ has positive
application probability at any time.

\subsection{Attractors and state transition graph}

A general definition of an attractor is based on the state transition graph
$G=(B^n,T)$ having all state vectors $B^n$ as its node set. Directed arcs in
this graph are direct state transitions. So there is an arc $(x,y) \in T$ from
$x \in B^n$ to $y \in B^n$ if there is an operator $U \in \mathcal{U}$ such that
$Ux=y$ and $y \neq x$. The fixed points of the dynamics are exactly those
state vectors that do not have outgoing edges.

An {\em attractor} of the dynamics is a sink component of the state transition
graph. A non-empty set $S \subseteq B^n$ is a sink component of $(B^n,T)$ if the
following two properties are fulfilled.
\begin{itemize}
\item[(i)] For all arcs $(x,y) \in T$ with $x \in S$, also $y \in S$.
\item[(ii)] No proper non-empty subset of $S$ has property (i).
\end{itemize}
In plain words, $S$ is a minimal non-empty set of state vectors from which no
arcs point to nodes outside $S$. By $\mathcal{A}$ we denote the set of all
attractors of the system under consideration. Note that $\mathcal{A}$ is a set
of pairwise disjoint sets of state vectors.

\subsection{Reduced dynamics}

What happens with the set of attractors under small modifications of the
dynamics? Let us consider the case of adding a single arc $(x,y) \notin T$ to
the state transition graph. Then the modified state transition graph with arc
set $T\prime =T \cup \{(x,y)\}$ has an attractor set $A^\prime$ such that
\begin{enumerate}
\item $|\mathcal{A}| \ge |\mathcal{A^\prime}|$; and 
\item For all $R \in \mathcal{A^\prime}$ there is an $S \in A$ such that
$S \subseteq R$.
\end{enumerate}
In plain words, the addition of arcs to the state transition graph may reduce
but not increase the number of attractors. Each attractor of the augmented state
transition graph is contained in an original attractor. Conversely, the removal
of arcs may only increase the number of attractors; each attractor of the
original dynamics is contained in an attractor of the dynamics reduced by arc
removal. This insight suggests to systematically remove arcs from the state
transition graph to obtain a reduced dynamics in which the attractors are easier
to find. The found attractors can then be used as seeds for the search for the
attractors in the original dynamics. 

We perform the reduction of the dynamics at the level of the update operators.
Considering the single-node updates again, we define the $b$-retaining operator
$U_i^{(b)}$ for a node $i$ and a Boolean state $b \in \{0,1\}$ by
\begin{equation}
U_i^{(b)}x =\left\{ \begin{array}{rl}
x & \text{if } x_i=b \\
Ux & \text{otherwise}
\end{array} \right.
\end{equation}
If we replace the operator $U_i$ by the corresponding $b$-retaining operator,
node $i$ can no longer switch from state $b$ to state $1-b$. This causes the
removal of arcs $(x,y)$ with $x_i=b$ and $y_i=1-b$ from the state transition
graph while all other arcs remain unaffected. Consequently, additional
attractors may be obtained while the existing ones become smaller, as discussed
in the preceding paragraph. 

The goal is now to replace a suitably chosen subset of all update operators
by state-retaining operators such that each attractor shrinks to a single state
vector, a fixed point. This is certainly the case when the reduced state
transition graph does not have a directed cycle. Along such a cycle, the nodes
that change state must do so in both directions, flipping from 0 to 1 equally
often as from 1 to 0. On the other hand, these switching nodes induce a
subnetwork of the Boolean network containing a cycle. Then by contraposition, we
see how cycles in the reduced state transition graph can be suppressed: Each
cycle on the Boolean network must contain at least one node that changes state
in only one direction. So we choose a set of nodes $X$ that hits each cycle of
the graph at least once. Such a set is called a feedback vertex set. For each
node in $i \in X$, the update operator $U_i$ is replaced by a state-retaining
operator, ensuring that each cycle of the Boolean network contains a node that
does not flip in both directions in the reduced dynamics.

Since $X$ is a feedback vertex set of the Boolean network, the subnetwork
induced by the remaining nodes $Y=n_\rfloor \setminus X$ does not have cycles.
It is a feed-forward network. Therefore there is a topological sorting of $Y$:
The indices of the nodes in $Y$ can be permuted such that arcs between nodes in
$Y$ go only from a lower index to a higher index. Given that the system is
in a fixed point, the states of the nodes in $X$ are sufficient to calculate
the states of the remaining nodes $Y$ as well. This is useful for finding
all fixed points of the reduced dynamics: For each state vector on $X$,
complete it to a state vector on the whole network and check if this state
vector is a fixed point or not.
 
The set $P$ of all fixed points of the reduced dynamics serves as a starting
point for constructing the attractor set $\mathcal{A}$ of the original dynamics.
Each attractor $S \in \mathcal{A}$ contains at least one of the state vectors
in $P$. Thus for each $x^\ast \in P$ we calculate the set of all state vectors
reachable from the $x^\ast$. If this set contains another element of $P$
then $x^\ast$ may be discarded. Otherwise this set is an attractor.

\section{Method for finding attractors} \label{sec:method}

\begin{figure}
\centerline{\includegraphics[width=0.5\textwidth]{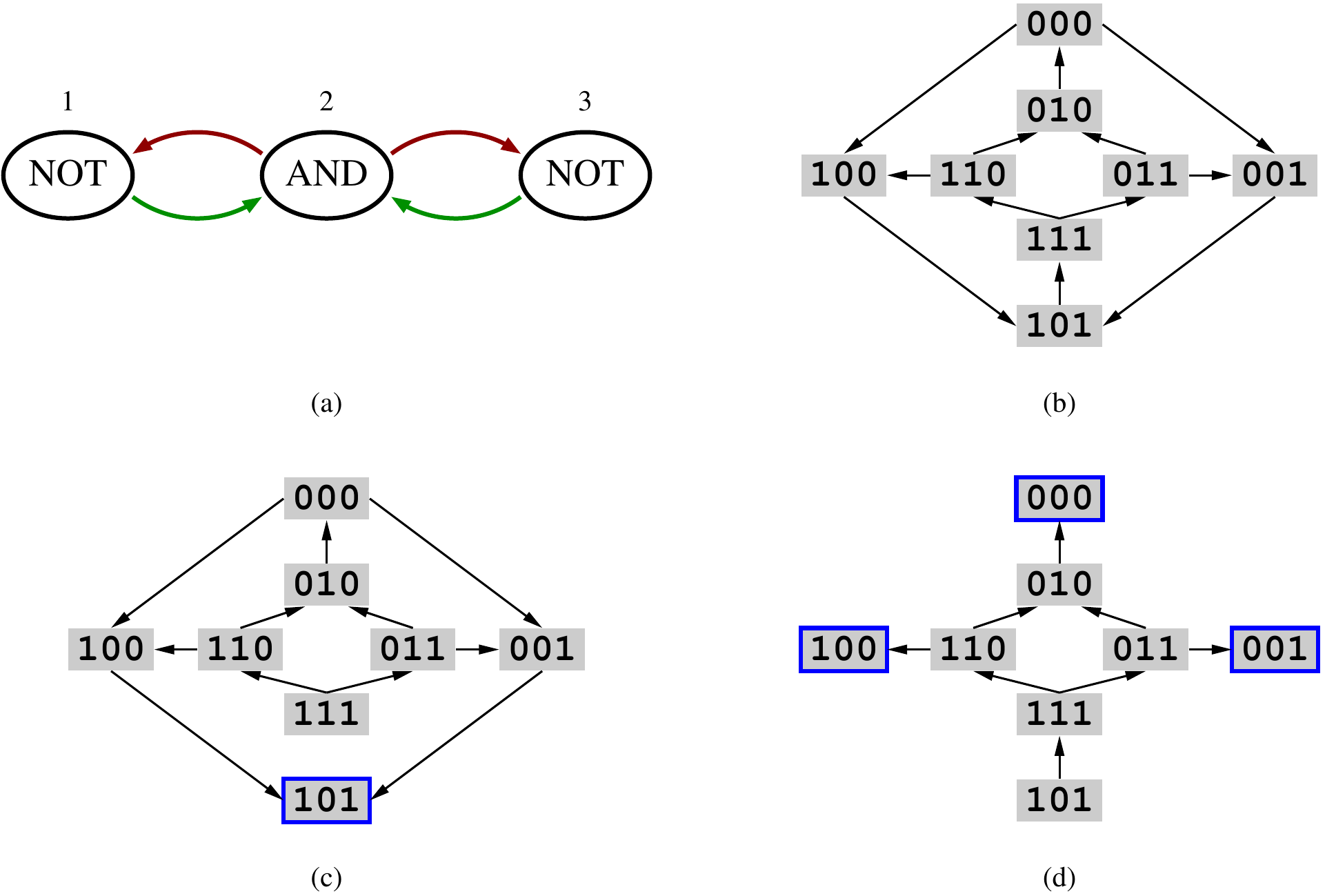}}
\caption{\label{fig:method_illu}
(a) A Boolean network with three nodes and (b) its state transition graph $G$
for asynchronous single node update. 
Since $G$ is strongly connected, there is a single attractor comprising all
states. The Boolean network may be made acyclic by removing node 2 or by
removing nodes 1 and 3. Thus $\{2\}$ and $\{ 1,3\}$ are the minimal feedback
vertex sets (FVS). Panel (c) is the reduced state transition graph for
FVS $\{2\}$ and $b_2=0$, where node $2$ cannot leave state $x_2=0$.
Network state $101$ is a fixed point of this reduced dynamics. Panel
(d) shows the reduced state transition graph for the larger FVS $\{1,3\}$ with
retained states $b_1=b_3=0$. Here the reduced dynamics has a set of 
three fixed points, $P=\{000, 001, 100\}$, inside the original attractor.}
\end{figure}

The computational method for finding attractors falls into three stages: (A)
Establish a feedback vertex set that defines which update operators are made
state-retaining. (B) Find the fixed point set $\mathcal{A^\ast}$ of the reduced
dynamics by enumeration (C) Traverse the original state transition graph
departing from the state vectors in $\mathcal{A^\ast}$.
A small example is illustrated in Figure~\ref{fig:method_illu}. 

\subsection{Feedback vertex set}

A feedback vertex set $X$ of the given Boolean network is determined as follows.
Initialize $X$ as the whole node set $n_\rfloor$. Loop: Draw a node $i \in X$ at
random (flat distribution). If $X \setminus \{i\}$ is a feedback vertex set, set
$X \setminus \{i\}$. Repeat this loop until $X$ does not have a proper subset
being a feedback vertex set. This very simple method tends to generate a small
but not necessarily globally minimal feedback vertex set. 

\subsection{Fixed points of reduced dynamics}

In addition to the feedback vertex set $X$, the retained state $b_i$
needs to be determined for each $i \in X$. Here we choose $b_i$ to be
the state that $f_i$ assumes for most values of the argument. If there is
a tie between 0s and 1s, $b_i$ is drawn at random from $B$ with equal
probabilities.

Without loss of generality we assume an indexing of the nodes such that $X
=\{1,\dots,m\}$ and $m+1,m+2,\dots,n$ is a topological sorting of the
feed-forward subnetwork induced by the remaining nodes $n_\rfloor \setminus X
=:Y$. After initializing the set $P$ of fixed points as the empty set, we
perform the following nested loops. The outer loop runs over all partial state
vectors $x \in B^m$. The first inner loop runs over the vertices in $i \in Y$ in
the order of topological sorting, calculating the  state
$x_i=f_i(x_1,\dots,x_{i-1})$. After finishing the first inner loop, a second
inner loop checks if
\begin{equation}
U_i^{(b_i)} x = x
\end{equation}
for all $i \in X$. If yes, $x$ is a fixed point of the reduced dynamics so $x$
is included in $P$. Otherwise $x$ is discarded.\\

\subsection{Original attractors}

After initializing the set of attractors $\mathcal{A}$ as the empty set,
attractor finding is performed as the following steps. (1) Draw an element
$x^\ast \in P$ and remove it from $P$. (2) Perform a depth first
search on the state transition graph starting at $x^\ast \in P$.
(3) If the search encounters an element $x \in P$, it terminates
immediately and the search result is discarded. Otherwise the search runs
until no further unvisited state vectors are found. Then the set $S$ of state
vectors visited during the search is included in $\mathcal{A}$ as an attractor.
(4) If $P$ is not empty, resume at (1), otherwise all attractors have been
found. Figure~\ref{fig:method_illu} shows a case where the reduced dynamics
has several fixed points contained in the same original attractor so search
results would be discarded in step (3).

\section{Results} \label{sec:results}

\begin{figure}
\centerline{\includegraphics[width=0.50\textwidth]
{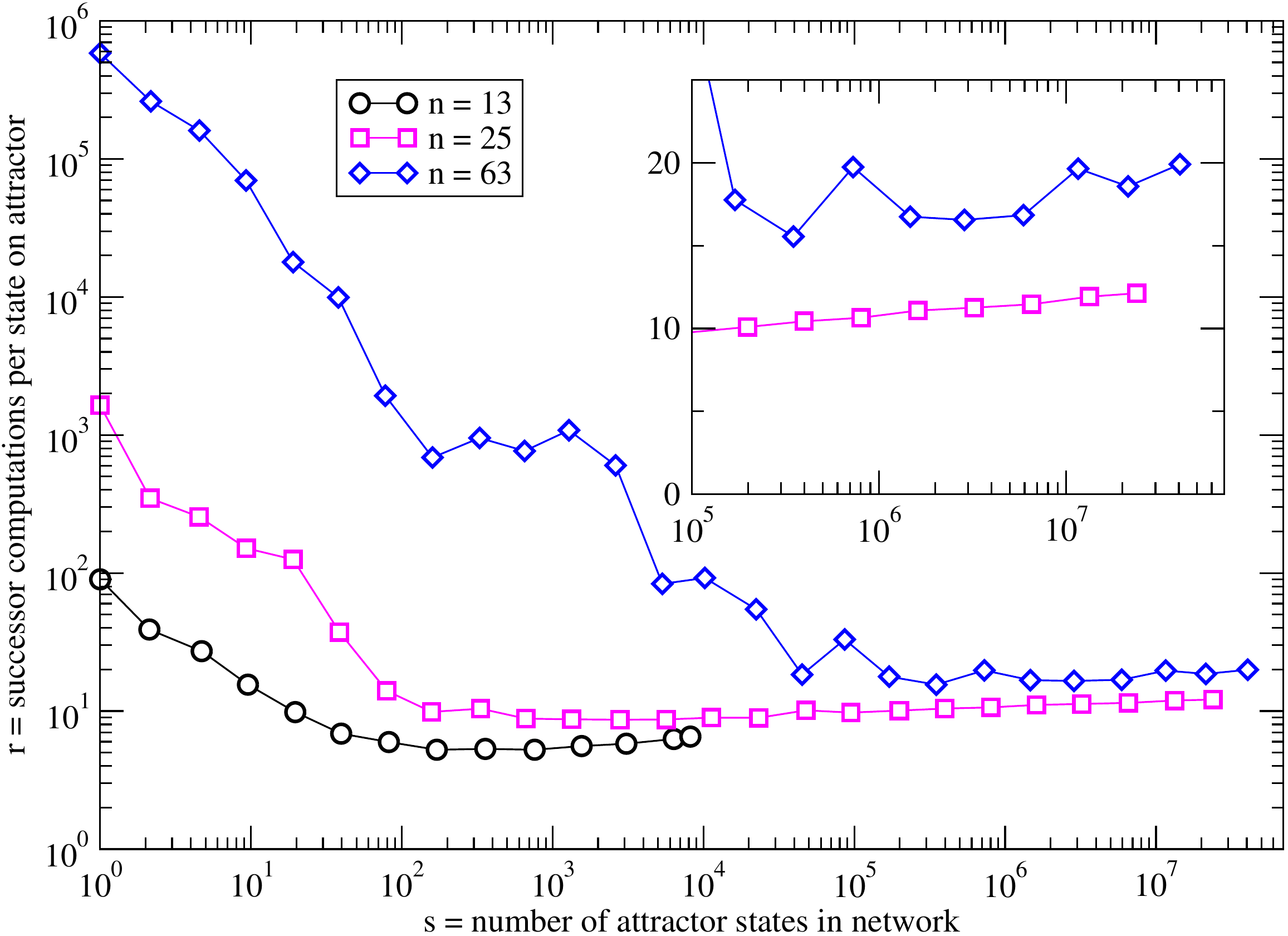}}
\caption{\label{fig:eff_ave}
Computational effort of attractor finding measured as the ratio $r$ between the number
of states $t$ calculated and the number of states $s$ actually contained in attractors.
The inset shows a linear-log plot of the same data pointing out the weak increase of
$r$ with $s$. The method is applied to Random Boolean networks with
$K=2$ inputs per node (critical Kauffman networks). Ensembles contain $10^4$
independently generated network instances for each system size $n \in \{13,25,63\}$.
A realization yields a data point $(s,r)$ with $s$ and $r$ as defined in the main text.
Each plotted point is an average over data points with $s \in [2^k,2^{k+1}-1]$,
$k=0,1,2,\dots$ (logarithmic binning). For $n=63$, only $8551$ data points are used.
On the remaining instances, the computation runs out of memory.}
\end{figure}

\begin{figure}
\centerline{\includegraphics[width=0.5\textwidth]
{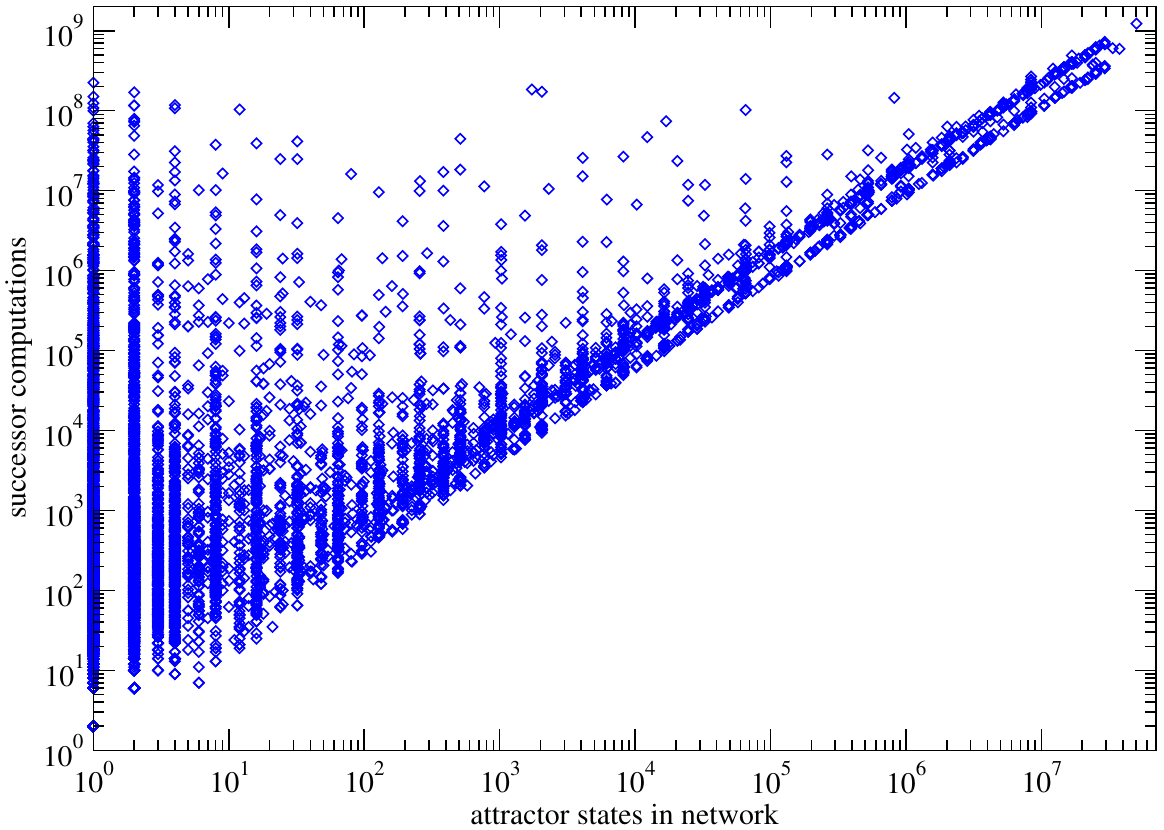}}
\caption{\label{fig:eff_scatter}
Scatter plot of computational steps $t$ versus number of states on attractors
$s$. Each data point represents a network realization with $n=63$ nodes. See caption
of \protect\ref{fig:eff_ave} for details.
}
\end{figure}

Standard random Boolean networks with $K=2$ inputs per node are generated as
test instances for the method as follows. Each node $i$ is assigned randomly
and independently one out of the 16 Boolean functions of two variables and a
pair of nodes from which node $i$ receives input.

We test the method with various system sizes up to $n=63$ and several
independent realizations of a network. As a first qualitative result we find that
the computational time of steps A and B of the method (finding $X$ and $P$) is
negligible compared with the final effort of searching the state transition
graph. Therefore we measure the computational cost in terms of the number $t$ of
successor states computed. Note that $t$ may be larger than $2^n$ because
a state with several predecessor states may be computed more than once.
A lower bound for $t$ is the number $s$ of state vectors
actually contained in attractors where
\begin{equation}
s= \sum_{S \in \mathcal A} |S|~.
\end{equation}
The method is efficient if the ratio $r:=t/s$ is small.
Figure~\ref{fig:eff_ave} shows that low values of $r$ are obtained for
intermediate values of $s$. With increasing $s$, the value of $r$ grows
moderately. The functional form might be logarithmic, $r \sim \log s$. More
extensive numerical simulations or analytical estimates are required to
clarify the growth. The scatter plot in Figure~\ref{fig:eff_scatter}
shows that fluctuations of $t$ become small for large $s$.

At network sizes $n\ge 30$, the computation does not succeed for all the
$10000$ instances because it exceeds the allocated memory. At $n=63$, a
fraction of $14.5 \%$ is unsuccessful because the depth first search of $G$
does not stay within a maximum depth of $3 \times 10^7$. Restarting with a
different random number for the $1449$ unsuccessful instances leads to a
complete computation in merely $19$ additional cases. This suggests that most
of the failing instances are intrinsically difficult and do not fail because of
an unfortunate choice of the feedback vertex set.

\section{Concluding remarks} \label{sec:con}

As a proof of concept we have introduced and used the method in its simplest
form. Several extensions of the method may help to increase the efficiency
and allow computation of attractors for larger and denser networks.

Smaller feedback vertex sets may be found by replacing our simple greedy
approach with a more advanced method. Having smaller sets $X$ tends to reduce
the set of fixed points $P$ of the reduced dynamics. It remains to be seen if
smaller $P$ leads to systematically shorter searches of the state transition
graph in the last phase of the computation.

One could attempt to explicitly guide depth first search towards other elements
in $P$ such that futile searches would terminate faster. The search then
should first choose as the next state vector a predecessor that reduces the
distance to another element in $P$. Alternatively, systematically different
graph traversals, e.g.\ {\em breadth first search}, might be tried out.

As the greatest limitation of the method, we experienced the need to store all
the state vectors of an attractor during the traversal of the state transition
graph. Therefore for some of the realizations at large $n$ the computation ran
out of memory. Space efficiency might be gained by a compression of sets of
state vectors during the traversal. Here it may help that attractors of sparse
networks often have a combinatorial product structure \cite{Albert:1999}.

Can the method be modified to work with synchronously updated Boolean systems
as well? Any synchronous system can be emulated with asynchronous
update under sufficient extension of the state space \cite{Nehaniv:2004}. It
would be interesting if the pruning of arcs in the state transition
graph is a general principle applicable to a larger class of discrete
dynamics.

\acknowledgments
The authors thank Fakhteh Ghanbarnejad for a careful reading of the draft. KK
acknowledges funding from VolkswagenStiftung. 

\bibliography{scoda}

\end{document}